# Prediction of helium vapor quality in steady state two-phase operation of SST-1 superconducting toroidal field magnets

G. K. Singh, R. Panchal, V.L. Tanna and S. Pradhan

Institute for Plasma Research, HBNI, Bhat, Gandhinagar-382 428 Gujarat, India

*Abstract* –**Steady State Superconducting Tokamak (SST-1) at the Institute for Plasma Research (IPR) is the first superconducting Tokamak in India and is an `operational device'. Superconducting Magnets System (SCMS) in SST-1 comprises of sixteen Toroidal field (TF) magnets and nine Poloidal Field (PF) magnets employing cable-in-conduit-conductor (CICC) of multi-filamentary high current carrying high field compatible multiply stabilized NbTi/Cu superconducting strands. SST-1 superconducting TF magnets are successfully and regularly operated in a cryo-stable manner being cooled with two-phase (TP) flow helium. The typical operating pressure of the TP helium is 1.6 bar (a) and the operating temperature is the corresponding saturation temperature. The SCMS cold mass is nearly thirty two tons and has a typical cool-down time of about 14 days from 300 K down to 4.5 K using helium refrigerator/liquefier (HRL) system of equivalent cooling capacity of 1350 W at 4.5 K. Using the available experimental data from the HRL, we have estimated the vapor quality during the cryo-stable operation of the TF magnets using the well-known correlation of two-phase flow. In this paper, we report the detailed characteristics of two-phase flow for given thermo-hydraulic conditions during long steady state operation of the SST-1 TF magnets as observed in the SST-1 experimental campaigns.**

## I. INTRODUCTION

SST-1 is working device at the Institute for Plasma Research, configured for steady state plasma experiments and to validate the advance tokamak technologies [1-2]. SST-1 superconducting magnet system (SCMS) comprises of sixteen superconducting Toroidal Field (TF) magnets and nine Poloidal Field (PF) magnets. The SST-1 TF cable-in-conduit-conductor (CICC) consists of 135 numbers of NbTi/Cu matrix stands of 0.85 mm diameter in a 3 x 3 x 3 x 5 cabling pattern tightly compacted in a stainless steel jacket of outer square dimensions 14.8 x 14.8 mm$^2$ having a conduit thickness of 1.5 mm [3]. This stainless steel jacket provides rigidity to strands against mechanical disturbances and also acts as narrow cryostat channel for the cooling of these twisted and bundled strands. The typical helium void fraction of this CICC is about ~ 40% [4-5]. Figure 1 shows the cross-sectional view of typical CICC of SST-1 TF magnets. The TF magnets in SST-1 have been wound in double pancake configuration with twelve parallel hydraulic paths, each having a hydraulic length of ~ 48 m. The total forced-flow two phase helium flow rate for all these 192 parallel paths of 16 magnets is ~ 60 g/s at supply pressure of 1.6 bar (a) and return pressure at 1.4 bar (a) in the cold conditions. In this paper, we have investigated the vapor quality in the steady state operation for the SST-1 TF magnets.

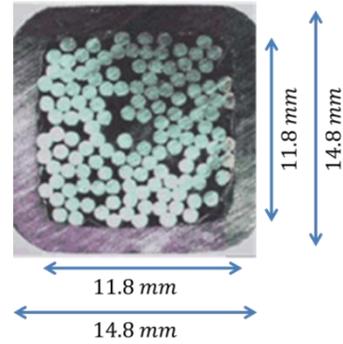

Fig.1. Cross-section of a typical SST-1 CICC [4]

The SST-1 TF magnets are cooled down in a controlled manner maintaining a temperature difference of < 50 K between the maximum and minimum temperature anywhere on the magnets surface in order to avoid any undue thermal stress. The cold helium flows through the void space given in the CICC. In order to understand the flow behaviour inside the complex geometry of CICC, one encounters difficulty in analysing the complex thermo-hydraulic characteristics especially when two-phase cooling is employed.

Viscous pressure drop across magnets and the static heat load acting on the magnet vaporises helium flowing inside magnet. Thus, a fraction of the liquid helium gets vaporised resulting in two phase (TP) flow of helium of certain flow regime in the cooling channel of the CICC. It is very difficult to measure mass flow rate and other hydraulic parameters under the two phase flow condition. The mass flow rate measured (by an `orifice meter') at the inlet of the SST-1 TF magnets is purely single phase (liquid) since the liquid is passed through a sub-cooler Dewar heat exchanger prior to its entry into the SST-1 TF magnets. Considering that fluid is in pure liquid phase at the inlet, the vapor quality evolution at the outlet ($x_o$) is critical to know. The problem is thus more complicated as the density and viscosity of the two phase mixture cannot be predicted without knowing the vapor void ($\alpha$). Knowledge of the vapor quality factor parameter greatly helps to simplify the two phase problem. In order to estimate quality at the outlet ($x_o$), we used Lockhart-Martinelli Correlation [10-12]; through which the quality can be predicted from the experiment data with certain assumptions and equivalent heat load can be estimated using heat balance equation. Toroidal field magnets specifications in SST-1 magnet system are listed in Table 1 below,

Table-1: TF Magnet (CICC) Specifications

|  | Unit | Value |
| --- | --- | --- |
| No. of magnets |  | 16 |
| Path per magnet |  | 12 |
| No of Paths |  | 192 |
| Each path length | m | 48 |
| Outer Dimension($L_{CICC}$) | m | 1.48E-02 |
| Inner Dimension ($l_{CICC}$) | m | 1.18E-02 |
| Diameter of Strand, $D_{st}$ | m | 8.60E-04 |
| Total Area ,$A_t$ | m2 | 1.39E-04 |
| Flow area of LHe, $A_{he}$ | m2 | 6.08E-05 |
| Wetted Perimeter, $P_{cool}$ | m | 3.14E-01 |
| Hydraulic Diameter, $D_h$ | m | 7.75E-04 |

The PF magnets are installed in the vicinity of the TF magnets and they have thermal contacts with TF magnets. Before achieving the cryo stable conditions in the TF magnets, we stopped cold helium flow to the PF magnets at about 24 K at the outlet of the PF magnets and allowed them to rise their temperatures over few days depending upon their cold masses and heat capacity.

## II. TF MAGNET HYDRAULIC BEHAVIOUR UNDER SINGLE PHASE

As discussed earlier, each TF magnet is wound with six double pancakes consisting of twelve equal and parallel hydraulic paths of 48 m each. There are such sixteen similar TF magnets in SST-1. The pressure drop across the TF magnets are more or less same (within 8-10% variation amongst the magnets largely because of the variations in the void fractions) as that of the individual path length. Therefore, the flow rate is assumed to be uniform and equally distributed in all paths of TF magnet. Starting from 300 K to 80 K, the typical cool-down rate is about 1.0 K/h whereas in the range of 80 K – 4.5 K, it is typically 0.5-0.6 K/h. With these cool down rates, it took about 14 days for cool down and to achieve cryo stable conditions within the TF magnets (~5.0 K). Figures 2 and 3 represent the single phase mass flow rate for single channel CICC path and pressure drop characteristics of during TF magnets cool-down process.

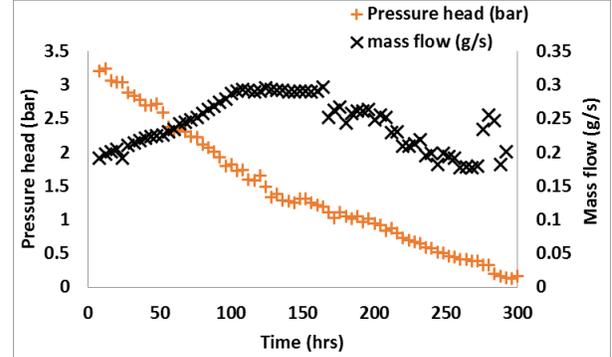

Fig.2. Mass flow and pressure head required for cool down of TF magnets

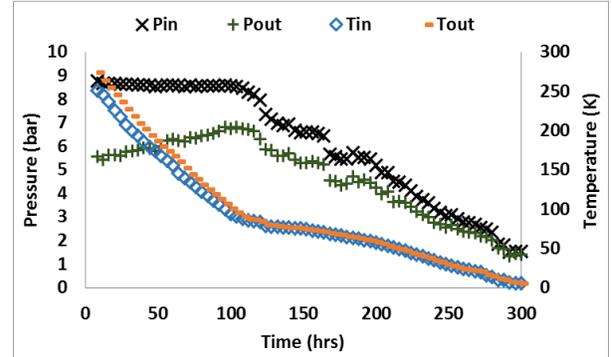

Fig.3. Inlet and outlet pressure-temperature variation for cool down of magnets

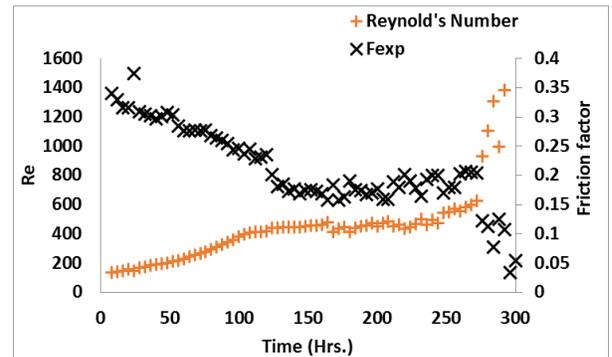

Fig.4. Reynolds number and experiment friction factor characteristics during cool down of magnets

During the cooling-down process, as the temperature drops, helium gas viscosity decreases and density consistently increases. This helps the cooling down process to progress fast. During the cool-down process, as helium density increases, mass flow rate increases, thereby Reynolds number increases and the friction factor decreases. The relationship between the Re and f are shown as a function of cool-down in Figure 4.An increase in the Reynolds number, "Re" results in the reduction of the friction factor. Accordingly pressure head requirement reduces as has been observed during the experiments in Figure 2. Superconductivity is achieved below 9.0 K after which the SST-1 TF magnets are operated in a steady state two phase flow of liquid helium. The Reynolds number and friction factor is estimated using equation 1-2 given below.

$$Re = \frac{4.\dot{m}}{\eta.P_{cool}} \qquad (1)$$

$$f = \frac{2.\Delta P.\rho.D_h.A_{he}^2}{L.\dot{m}^2} \qquad (2)$$

$Re$ =Reynolds number
$\dot{m}$ =Mass flow rate of coolant
$\Delta P$ =Pressure drop across magnet
$L$ =Path Length

## III.    TWO PHASE FLOW AND VAPOR QUALITY ESTIMATION

After achieving the cool down of the TF magnets, we obtain steady state cryo stable two phase helium flow conditions within the TF magnets prior to charging the magnets. It is observed that during the steady state two phase conditions, the outlet temperature is less than that of the inlet temperature [6] as shown in Figure 5 (higher readings are due to external heat load on sensor) and the pressure drop in case of the two phase flow is greater than that of the single phase flow for the same mass flow. As the heat load of the system increases, for a given mass flow rate the quality value increases. As compared to the single phase, there is large pressure drop accounted for the two phase flow conditions. At a given fixed supply pressure, one observers reduction in the pressure at the outlet. Under the two phase dome, there is a specific fixed saturation temperature corresponding to the pressure at the outlet. Thereby, we observed the reduction of the temperature at the outlet as compared to the boiling temperature at the supply side. Additionally, under the two phase flow condition, we need to consider other parameters viz. flow quality (x), void fraction (α) and slip ratio (S) etc. These parameters are generally given as follows [7-9]:

$$\alpha = \frac{1}{1 + \left(\left(\frac{\rho_V}{\rho_L}\right).S.\frac{(1-x)}{x}\right)} \qquad (3)$$

$$\alpha = \frac{A_V}{(A_V + A_L)} \qquad (4)$$

$$x = \frac{\dot{m}_V}{(\dot{m}_V + \dot{m}_L)} \qquad (5)$$

$$S = \frac{v_V}{v_L} \qquad (6)$$

$$\rho_m = \alpha\rho_V + (1-\alpha)\rho_L \qquad (7)$$

$$\eta_m = \alpha\eta_V + (1-\alpha)\eta_L \qquad (8)$$

$\rho_V$ =Vapor density (kg/m$^3$)
$\rho_L$ =Liquid density (kg/m$^3$)
$A_V$ =Vapor area (m$^2$)
$A_L$ =Liquid area (m$^2$)
$\dot{m}_V$ =Mass flow of vapor (kg/s)
$\dot{m}_L$ =Mass flow of liquid (kg/s)
$\rho_m$ =Mixture density (kg/m$^3$)
$\eta_m$ =Mixture viscosity (Pa*s)

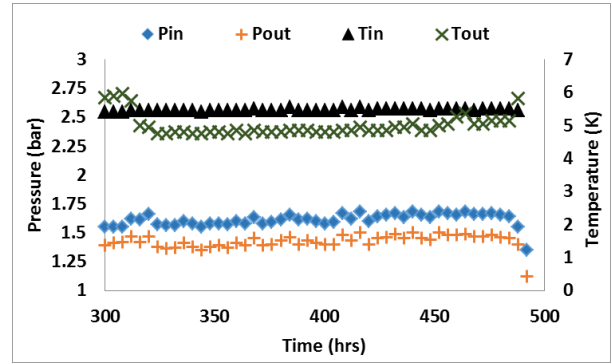

Fig.5. Inlet and outlet pressure-temperature variation in steady two phase flow condition

The range of x, α varies from 0 to 1. The mass flow rate, mixture density and viscosity are very difficult to measure in two phase conditions when vapor quality is unknown. Using equation 3, the vapor void can be calculated if vapor quality is known for S=1, and hence mixture density and viscosity is calculated using equation 7-8. Using the Lockhart Martinelli correlation [10-12], we have estimated vapor quality from experiment, and for this we have used  single phase pressure drop at a cold temperature and the two phase pressure drop in steady state conditions, inlet and outlet (pressure, temperature) conditions for saturated helium properties.

In equation 3, we assume homogeneous flow, i.e. the vapor velocity is equal to liquid velocity ($v_V = v_L$) in two phase flow. With this assumption the slip ratio(S) = 1. Thus, void fraction now depends upon the vapor quality and density variation between the two phases of helium at a particular pressure and temperature. Using the above assumptions as well as separated flow condition and a pure liquid in the inlet of TF magnet

$(x_i = 0)$, we have estimated the vapor quality using Lockhart-Martinelli correlation as below:

*Lockhart-Martinelli Correlation:*

Lockhart-Martinelli Correlation [10-12] (equation 9-13) is used to estimate two phase pressure drop using the single phase pressure drop and the saturated properties of the fluid. Using this correlation, one can also estimate vapor quality if two phase and single phase pressure drops and saturated properties of helium is known.

For homogenous flow (equation 9-10):

$$\frac{\Delta P_{tp}}{\Delta P_L} = \frac{1}{(x_o - x_i)} \int_{x_i}^{x_o} \emptyset_L{}^2 dx \qquad (9)$$

$$\emptyset_{LO}{}^2 = \left\{1 + x\left(\frac{\rho_L}{\rho_V} - 1\right)\right\}\left\{1 + x\left(\frac{\eta_L}{\eta_V} - 1\right)\right\}^{-0.25} \qquad (10)$$

Using Binomial expansion in second term of equation (10), under the assumption $\left\{x\left(\frac{\eta_L}{\eta_V} - 1\right) < 1\right\}$ can be written as;

$$\emptyset_L{}^2 = \left\{1 + x\left(\frac{\rho_L}{\rho_V} - 1\right)\right\}\left\{1 - \frac{x}{4}\left(\frac{\eta_L}{\eta_V} - 1\right)\right\}$$

For separated flow (equation 10-13):

$$\frac{\Delta P_{tp}}{\Delta P_V} = \emptyset_{vtt}^2 \qquad (11)$$

$$\emptyset_{vtt}{}^2 = 1 + CX_{tt} + X_{tt}^2 \text{ , for } Re_L < 4000 \qquad (12)$$

$$X_{tt}^2 = \left(\frac{1-x}{x}\right)^{0.9}\left(\frac{\rho_v}{\rho_L}\right)^{0.5}\left(\frac{\eta_L}{\eta_v}\right)^{0.1} \qquad (13)$$

Where C=5 for laminar flow.

$\Delta P_{tp}$ =Two phase pressure drop (Pa)
$\Delta P_v$ =Single phase vapor pressure drop (Pa)
$\Delta P_L$ =Single phase Liquid pressure drop (Pa)
$\emptyset_{LO}{}^2, \emptyset_{gtt}{}^2$ =Two-phase multiplier
$\eta_L$ =Liquid viscosity (Pa-s)
$\eta_V$ =Vapor viscosity (Pa-s)

Under steady two phase helium condition, the required mass flow rate is maintained in order to get sufficient amount of liquid helium at the outlet and maintain the cryo stability of the superconducting magnets. The inlet of the TF magnets are considered to be saturated liquid helium $(x_i = 0)$ and the two-phase pressure drop in the magnets is obtained through the SST-1 campaign experimental data. The single phase pressure drop is also obtained during the cool down. The outlet vapor qualities are iterated using the Lockhart-Martinelli correlation.

The variation in the two phase flow pressure drop shows that the vapor quality is varied based on heat load and it can be observed in the predicted vapor quality profile (Figure 6). The two phase multiplier is a function of density and viscosity of the liquid and the vapor phase of helium, obtained from the saturated properties of liquid helium. The vapor quality predicted using this method is shown in Figure 6.

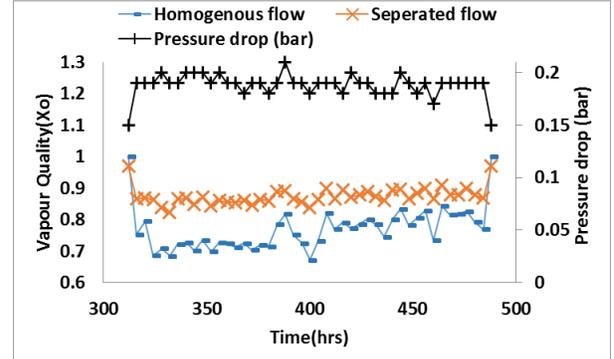

## IV. RESULTS AND DISCUSSIONS

We have estimated vapor quality and heat load on TF magnets and it is found to be in good agreement with Helium plant cooling capacity as shown in Table 2. Figures 6 shows, the pressure drop relation with vapor quality, and when the mass flow rate is increased to maintain the operating condition of the magnets reduction in vapor quality is observed. During SST-1 experiments, if the mass flow rate at a particular heat load is unable to maintain two phase conditions in the magnet, then the outlet temperature starts to rise. In order to maintain steady state two phase flows, the mass flow rate is usually increased using the control valve and a requisite mass flow is maintained for steady magnet operation, resulting in pressure drop increase and hence quality decreasing.

Figure 7 shows, average vapor quality and its variation over the days of operation of TF coils. Initially vapour quality improves as PF coils are bypassed and mass flow is increased in TF magnets to achieve Cryo-stability. Over the days, PF coils impose heat load on TF coils and due to which vapor quality is observed to be increased.

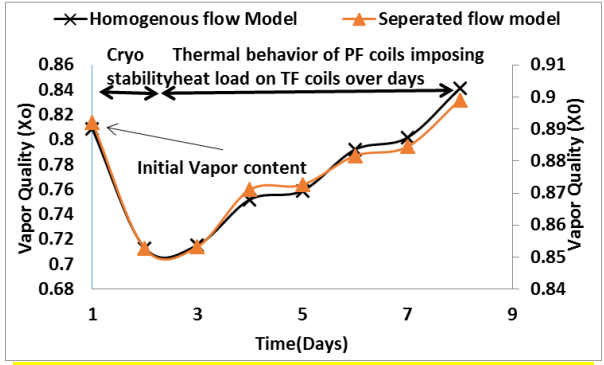

In Table 2, we have repeated this study for the 18th and 19th SST-1 campaign, and we have observed that the results are in agreement with those of the 17th campaign; therefore, it is a bench marking our analytical tool towards validation. We have shown in Table-2, the predicted vapor quality and corresponding heat load (using equation 14) based on the best achieved data in the SST-1 experimental campaigns 17, 18 and 19. We predicted the average heat load of 822 W using homogeneous flow model and 936 W (approx. 1000 W) using separated flow model for the TF magnets using the experimental database and heat balance method under the specific condition where the PF magnets flow were stopped at about 24 K at the outlet and over days, they were allowed to rise their temperatures. During operation approximately 350 W headload is utilized to collect liquid helium in main control Dewar).

$$Q = \dot{m} L_v x_o \qquad (14)$$

Q = Heat Load (watts or J/s)

$\dot{m}$ = Mass flow rate (g/s)

$L_v$ = Latent heat of Vaporization (J/g)

Table 2: Vapor Quality and heat load is estimated using the best achieved data in the campaigns 17, 18 and 19 for the TF magnets

| SST-1 Campaign | $\dot{m}$ g/s | $P_{in}$ bar | $P_{out}$ bar | $X_0$ | Q(W) | $X_0$ | Q(W) |
|---|---|---|---|---|---|---|---|
| | | | | Homogeneous flow | | Separated flow | |
| 17th | 62 | 1.61 | 1.43 | 0.75 | 832 | 0.86 | 954 |
| 18th | 60 | 1.61 | 1.43 | 0.75 | 805 | 0.85 | 912 |
| 19th | 65 | 1.58 | 1.58 | 0.77 | 830 | 0.87 | 940 |

## V. CONCLUSION

The pressure head and quality factor analysis have been carried for the SST-1 TF magnets. In this paper, using Lockhart Martinelli relation and actual experimental data of the SST-1 TF magnets, we have tried to estimate the vapor quality. The results also shows that as the mass flow rate increases, corresponding pressure drop increases and hence the vapor quality decreases for a given heat load. Over the days due to heat load by PF coils increase in vapor quality is observed. This can be used as one of the efficient tool for analysing the two phase flow characteristics in complex flow geometry like CICC.

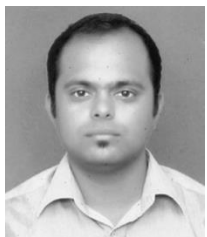

**Gaurav Kumar Singh** received M.Tech degree in nuclear engineering from the Pandit Deendayal Petroleum University, Gandhinagar, India, in 2014. He is currently with the Institute for Plasma Research, Gandhinagar, India, working as a research scholar. His research interests are study the physics of two-phase flows of cryogens, development of two phase flow sensor.

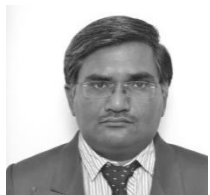

**Rohitkumar N. Panchal** received B.E. degree in Instrumentation and Control from the Nirma Institute of Technology, Ahmedabad, India, in 1999. He is currently an Instrument and control Engineer with the Cryogenics Division of Steady State Tokamak-1, Institute for Plasma Research, Gandhinagar. He works on the operation, control and maintenance of Helium Refrigeration and Liquefaction plant and its sub system. His research interest in Flow measurement especially for cryogenic.

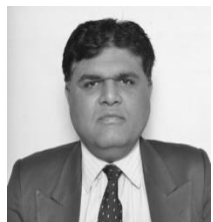

**Vipul L. Tanna** received the M.Sc. degree in physics from Saurashtra University, Rajkot, India, in 1992, and the Ph.D. degree from the University of Karlsruhe, Karlsruhe, Germany in 2006. He is currently working as the Division Head of SST-1 Cryogenics Division with the Institute for Plasma Research, Gandhinagar, India. His research interests include fusion-machine-relevant cryogenics, design and testing of large magnets, thermohydraulic simulation of magnets.

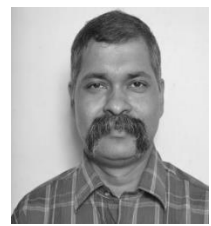

**Subrata Pradhan** received M.Sc. degree in physics from the Indian Institute of Technology, Kanpur, India, in 1989, and the Ph.D. degree in physics from Gujarat University, Ahmedabad, India, in 2004. He is currently with the Institute for Plasma Research, Gandhinagar, India, leading the Steady State Tokamak Program as Mission Leader. His research interests are the technology and engineering aspects of fusion-grade superconducting magnets, development of highly homogeneous superconducting magnets, and technologies related to the magnetically confined fusion devices, such as Tokamaks.